# Central Dogma Cycle and Network: A Model for Cell Memory

*Martin R. Schiller*[1,2]

[1]Nevada Institute of Personalized Medicine and School of Life Sciences, University of Nevada, 4505 Maryland Pkwy, Las Vegas, Nevada 89154-4004
[2]Heligenics Inc., 10530 Discovery Dr., Las Vegas, Nevada 89135

**Subtitle**
Central Dogma Cycle

**Abstract**
This paper proposes an extension of the traditional Central Dogma of molecular biology to a more dynamic model termed the Central Dogma Cycle (CDC) and a broader network called the Central Dogma Cyclic Network (CDCN). While the Central Dogma is necessary for genetic information flow, it is not sufficient to fully explain cellular memory and information management. The CDC incorporates additional well-established steps, including protein folding and protein networking, highlighting the cyclical nature of information flow in cells. This cyclic architecture is proposed as a key mechanism for cellular memory, drawing analogies to memory functions in computers, such as input, read, write, execute, and erase. The interconnected cycles within the CDCN, including metabolic cycles and signaling pathways, are suggested to function akin to latches in computer memory, contributing to the storage and processing of cellular information beyond nucleic acid sequences. Understanding cellular memory through this cyclic network model offers a new perspective on heredity, cell processes, and the potential disruptions in disease pathology.

## Introduction

The central dogma of molecular biology describes the flow of genetic information, whereby DNA self-replicates based on its sequence complementarity, is transcribed into a messenger RNA (mRNA) copy of the DNA template, and the ribosome translates the mRNA into a protein chain (**Fig. 1A**). Originally, proposed by Francis Crick in 1958 and later revised by Tom Watson in 1965, this Central Dogma model has since been revisited several times.[1,2,3–7] Recent advancements in multi-omics technologies have deepened our understanding of cell processes, revealing additional layers of complexity not captured in the Central Dogma. The Central Dogma is necessary for the flow of genetic information, but is not sufficient to pass on the information to new cells or even use it in the same cell. Thus, the Central Dogma does not completely enumerate the flow of genetic information. Furthermore, the Central Dogma does not include epigenetic control of transcription, or how protein complexes produce cell processes that drive the genetic flow of information. Herein, I propose an extension of the Central Dogma to a more dynamic model: The Central Dogma Cycle (CDC) and a broader network called the Central Dogma Cyclic Network (CDCN).

## Results

**The Central Dogma Cycle.** Recognizing that the Central Dogma is not merely a linear pathway but a more complex set of interrelated cycles, I introduce the concept of the Central Dogma Cycle (CDC) as an architecture for the majority of information flow in cells (**Fig. 1B**). The Central Dogma can be conceptually modified and extended with additional well-established steps into the CDC.

The appreciation of the role of epigenetic regulation in controlling transcription has become much better appreciated since the discovery of the Central Dogma.

After translation of a protein chain, the next step is protein folding. Previous studies of Christen Anfinsen's and his colleagues demonstrated that protein sequences contains the information necessary and sufficient for to fold a protein in a metastable three dimensional (3D) structure[8]. This information is independent of nucleic acids, as a protein can fold *in vitro*. Predicting proteins folds has been a longstanding scientific problem. Although homology modeling was the first widely adopted method for predicting protein folding, it was limited to proteins having high sequence similarity with a protein of established fold. The recent advent of machine learning approach with Alphafold 2 has significantly improved the broader predicative accuracy[9–11]. Despite these advancements and the knowledge that folding is driven by energetics, a cipher, code, or precise rules governing protein folding remains elusive. Notably, protein folding can also be considered an additional form of inherited memory, as its sequence determines the fold, and the protein sequence is a transformation of the DNA sequence of a gene.

In the next step of the CDC, folded proteins encode numerous types of specific biological functions such as enzymes, binding proteins, and structural proteins. The protein fold determines function. However, the emergence of function is a major challenge. While long-term experimental approaches are the gold standard, predictive models of function are in their infancy. David Baker's work on *de novo* design of proteins using machine learning for is cutting edge. However, there is little theoretical framework to predict the function of a protein from its sequence alone.[12–14] The best current approach is identifying structurally and functionally related domains by searching for homologous protein sequences.[12,15] While similar domains often exhibit similar molecular functions, this is not always the case. Notably, like a fold, the function can also be considered a memory that is inherited via the folded protein.

Predicting the emergence of protein function from sequence might benefit from recent studies from my laboratory and Heligenics Inc.[16–18] We have explored how intragenic epistasis (the dependent interactions of two or more amino acids in a protein) influences the Tat transcription factor. We demonstrated that all variants interact in specific ways that can be codified using digital logic gates to define relationships. These findings infer interaction of amino acids in protein may be based on circuits or networks of digital logics gates creating computational logic, akin to how logic gates are combined to generate functions in electrical engineering.[17,19,20] In this paper, we proposed extending this digital logic framework to broader aspects of protein biology and cellular function. This observation was the seed for the concept of the CDC presented herein (**Fig. 1B**).

In the final step of the CDC, proteins with specific functions, network with other proteins, macromolecules, and metabolites, producing emergent cell processes that control cell behavior. It is well established that proteins networks have many emergent properties such as those collected in the Gene Ontology database[21]. These emergent properties of the complexes include the processes that underlie steps in the CDC: DNA replication, epigenetic regulation, transcription, translation, protein folding, and protein networking. Each of these processes connect to a specific CDC step and close a directed cycle. Importantly, these processes also provide feedback regulation throughout the CDC, reinforcing a cyclical model of information flow, that like the central dogma, is also conserved in all living systems (**Fig. 1B**).

By framing genetic information flow as a cycle rather than a one-directional pathway, the CDC captures the intricate interdependencies that underlie cellular function. This perspective not only integrates many discoveries in molecular biology and omics, but also provides a more comprehensive framework and fresh interpretation for future research. As will become apparent, the cyclic nature of the CDC is likely related to cell memory.

**Comparing memory across humans, computers, and cells.** I digress to discuss fundamentals of memory. The Central Dogma is the primary mechanism of heredity in cells. Heredity is a form memory, transmitting genetic information from one cell to another, and from parent organisms to progeny by replicating its DNA. Heredity is not traditionally called "memory", which is more commonly associated with a brain or computer. Exploring the analogies of memory, provides new insights into inheritance and information flow in cells. Cells inherit and utilize many different functions, pathways, and processes not captured in the Central Dogma or the aforementioned CDC. For example, a cell has an inherited memory of how to control ion gradients.

There are five general memory functions observed among the nervous system, computers, and cells: input, read, write, execute, and erase, although there are semantic differences (**Table 1**). Additional common functions are combinations of these base functions. For example, copy combines read and write; while move combines reading with writing to a new location. Each system has memory that is classified into multiple types such as short-term and long-term. Although the architectures differ among systems, the fundamental memory functions are conserved. Memory involves reading an input and writing it to a location for storage. The memory can then be read, moved, executed with a function, or erased.

**Table 1: Memory functions across human nervous system, computers, and cells.**

| Privilege | Nervous System | Computers | Cells |
| --- | --- | --- | --- |
| *Input* | Sensory input | Device input (keyboard, mouse, sensors) | Protein factors, Environmental conditions, Reproduction |
| *Read* (access, view) | Short-term memory (hippocampus) | Read-only memory (ROM), Cache, RAM. and Hard Drive | DNA, chromatin, mRNA, Post-translational modifications (PTMs), epigenetic marks |
| *Write* (store, modify) | Long-term memory, neuroplasticity, synaptic changes | External Memory (Hard Drive), Cache, RAM, Hard Drive | DNA, mRNA, proteins, PTMs, epigenetic marks |
| *Execute* (move, duplicate, move, copy) | Memory consolidation synaptic plasticity, gene expression | Processor, Cache, RAM | Signaling pathways / trafficking, replication, transcription, translation |
| Erase (delete) | Short term memory loss, forgetting | Deletion from Hard Drive, RAM clearance | Apoptosis, PTMs, epigenetic modifications |

The *Input* function receives information from an external source. Humans receive sensory input by sight, hearing, smell, taste, and touch. Computers receive information from devices like keyboards and sensors. Cells take inputs from external primary messengers or environmental conditions. In all cases, the inputs are transmitted and written to a memory storage location.

The *Read* function, also called access or view, accesses stored information without modifying it. In the nervous system, this occurs first in sensory nerves, then in the hippocampus (short-term memory), and later from brain regions (long-term memory). In computers, information is read from input devices or storage locations, and written to Cache, RAM, and hard drives. In cells, genetic information is read from DNA and written by transcription and replication. Other forms of cellular memory have specialized reading, moving, executing, and writing collectively called cell processes.

The *Write* function, also somewhat synonymous with store or modify, reads existing memories as an input, and *writes* the information to a new location (stores) or overwrites information at an existing location (modify). In humans, sensory or environmental input are transmitted and stored in the hippocampus which is a read-write function, similar to computer RAM. This is considered short-term memory. Long-term storage involves synaptic plasticity and gene expression in other brain regions. Computers write data to RAM before committing it to external storage. In cells, inputs are written in many different biochemical ways. The cell has evolved multiple mechanisms for short-term storage such as chemical concentration gradients, capacitance, protein localization, post-translational modifications to proteins, covalent and structural modifications of chromatin (epigenetic), and gene expression. Longer term storage is in the form of DNA.

The *Execute* function reads stored information, processes it, and writes the output to a location. In computers, the processor loads instructions from RAM and carries them out. In the nervous system, short term memories stored in the Hippocampus are executed directly, or are processed and consolidated, writing them to other brains regions for long term memory. Cells are much more heterogeneous in how memory is processed. Cells execute many processes such as DNA replication, transcription, translation, or many others.

The *Erase* function, also called delete, removes information stored at a specific location. In the nervous systems, short-term memory in the hippocampus is purged upon consolidation to select for more persistent, longer-term memories in other brain regions. Likewise, in computer systems, short-term memory in RAM, or Cache in computers are often purged and select data is stored in hard drives, but unlike brains, can be selectively deleted. Cells routinely degrade mRNA, modify chromatin, or degrade proteins. Long-term memory in the form of DNA is generally not erased, similar to the nervous system.

By comparing memory across biological, computational, and neurological systems, we gain new insights into inheritance, information processing, and realize the broader concept of the memory in cells.

**Do cycles create memory in Biological Systems?** I question" How does a cell preserves memories that are not encoded in a nucleic acid sequence?" This question is addressed by considering an analogy to computer memory. When a cyclic circuit has an input that changes the output in response to a second input, this is an expression of memory. In computers, the fundamental unit of memory is a *bit,* which can be represented by a "*SR Latch*", a flip-flop two-component circuit with two NOR gates and two switch inputs (**Fig. 2B**).[22] Similarly, a "D latch" is a one input form of memory adding a simple circuit input to the SR-latch. Computer memory is built from clusters, usually *8* *bits* combining them to form a *"byte"*, which can be sequentially combined to produce larger memory blocks.

I propose that cells are using a similar cycle architecture to create and organize memories. In the 1930s, Albert Szent-Györgyi and others identified several key metabolic reactions that were eventually recognized to be part of the Tricarboxylic Acid cycle for oxidative metabolism, also called the Krebs cycle. The identification of the reaction of acetyl-CoA condensing with oxaloacetate to citrate by Krebs and Johnson closed the metabolic pathway reactions into a cycle.[23,24] In this cycle, Acetyl-Coenzyme A (CoA), Nicotinamide Adenine Dinucleotide (NAD$^+$), Guanosine diphosphate (GDP), Flavin adenine dinucleotide (FAD), and oxygen are inputs that drive the cycle and maintain the metabolic memory, which is passed from cell to cell and organisms to organism. Only the genes for the enzyme catalysts are passed by DNA.

While the Krebs cycle is one of the earliest discovered and most well-known biological cycles, cycles are pervasive features of most, if not all cellular and physiological processes. There are over 88,000 papers identified with the search term cycle in PubMed.[25] Metabolic cycles, in particular are widespread.

A cell can be modeled as a large set of interconnected cycles in which each cycle receives inputs and produces outputs to and from other cycles in the cell. There are many cycles that are already established. The Urea Cycle underlies nitrogen waste disposal and fatty acid synthesis and beta-oxidation form cycles that controls lipid metabolism.[26–28] The glyoxylate cycle is a variation of the Krebs cycle converting fats into sugars, the Calvin cycle is essential for carbon fixation for photosynthesis in plants and autotrophs. The centrosome cycle regulates its duplication, and the carbon cycle is an atmospheric cycle converting inorganic carbon gases into organic molecules.[29–32]

Additionally, energy-carrying molecules like NAD$^+$ and ATP are continuously regenerated in cycles (REDOX and the ATP cycles, respectively), forming the energetic drivers of many interconnected metabolic reactions.[33,34] Because ATP and NAD$^+$ drive numerous biochemical pathways, nearly all metabolic processes involving these molecules inherently function as cycles. Metabolic cycles contain a catabolic pathway that forms new molecule(s) and an anabolic pathway that breaking down molecules into their constituents.

Beyond metabolism, other biological cycles play critical roles in cellular and systemic function. The Cell Cycle regulates cell growth and division, and Circadian Rhythms govern biological clocks in nearly all living organisms.[35,36] Other cycles are the Secretory Pathway - a series of organelle cycles involved in protein and membrane trafficking and secretion, and neurotransmitter recycling ensuring efficient communication between neurons by reusing neurotransmitters in synaptic terminals; the synaptic vesicle cycle is part of this process.[37–39] Cycles can be identified by other names in the literature such as negative and positive feedback loops; e.g. endocrine feedback loops.[40,41] Also, any biomolecule or cellular entity such as the secretory pathway that undergoes synthesis and degradation, or constructive or destructive process into its building blocks operates within a homeostatic cycle, ensuring cellular balance and resource management. Phosphoinositide signaling in cells is an example.[42]

This perspective highlights the importance of cyclic processes in maintaining biological homeostasis and adaptability. Given the sheer number of interdependent cycles which likely numbers in the 10,000s, a cell can be viewed as a vast network of interconnected cycles, in which each cycle receives inputs and produces outputs that feed into other cycles.

Why has the cell adopted so many cycles as an organizing element? In metabolism, it was suggested that cycles help conserve metabolites. Another consideration is that many cycles have specific inputs and outputs, enabling them to regulate flow and energy through feedback.[43] However, I contend that the primary function of the cell's architecture of cycles is likely to serve as a form of memory, akin to how states are stored in in computer memory. Each cycle has multiple inputs and outputs, making them structurally similar to a *SR latch* or series of *latches* in computers.

Some evidence supports the encoding of cell memory outside of the inherited DNA: 1. Many cell processes continue in Red Blood Cells that lack nuclei and mitochondria, have a memory to perform such processes, and do not have DNA; 2. Most cell processes and their cycles are conserved among cells and organisms; 3. Protein can fold *in vitro* based on their sequence alone. 4. A protein fold holds a memory of molecular function. 5. Proteins, metabolites, and organelles are passed to cells during cell division; this also includes complete functioning cycles; and 6. cells have architectural designs similar to that of memory in a computer.

**Cells have an interconnected cycle architecture for memory, like RAM circuits in a computer.** The CDC was presented as a single cycle in **Fig. 1A**. However, the CDC is a much more complex series of interconnected cycles organized in a higher order cycle (**Fig. 2**).

Cell processes in the CDC (replication, epigenetic regulation, transcription, and translation, protein folding and function) emerged from a networked interaction of a collection of cell functions. The cell process that emerges from the network drive other cycles in the CDC. For example, mRNA from the transcription cycle enters the protein cycle and is translated into a protein by the ribosome. The cell process of translation is created by complexes of the protein and RNA products produced from other cycles in the CDC. This interconnected cycle architecture of the CDC acts like interconnected *bits*, similar to the connection of bits in a *byte* unit of computer memory.

The CDC cycles are related to memory functions. The DNA replication cycle is analogous to copying data for long-term storage on a computer hard drive. Access to the DNA for replication and transcription is controlled by an epigenetic cycle regulating modifications of the DNA and nucleosome. Epigenetics can be considered either short-term memory that is responsive to cell signaling or long-term memory for genetic information that can persist for generations. Transcription is similar to the copy function in a computer - a gene is transcribed into pre-mRNA and then processed into mature mRNAs, each of which has a variable half-life but can be considered short-term storage, akin to that in computer RAM, the short-term memory of transcription is referred to as gene expression, or collectively the transcriptome. Each cell type has a specific memory of which transcripts are expressed. Because most proteins have a half-life on the order of hours or days, they can be considered short-term memory like RAM in a computer, in which they are written and can be quickly accessed.[44,45]

Protein molecules reliably fold based on their sequence, thus having a form of memory that is added to the genetic memory proteins inherit from an mRNA sequence. Concerning the lifetime of a protein, the fold is a long-term memory, but short-term with respect to the DNA memory. Folding is executed, and the fold is written as it emerges from the ribosome during translation, and a correct fold is read through an emerged property, its specific function(s). Thus, a fold is a transformation of memory from the protein sequence and is analogous to storage in a hard drive.

A folded protein has a precise memory for a specific function(s). Some example functions are catalysis, light harvesting, binding small compounds, light harvesting, elasticity, interacting with other macromolecules, and many others. Since the function persists as long as the sequence and fold remain intact; the function should be considered a short-term memory, analogous to RAM. The DNA encoding the protein is a memory of the sequence, but not the fold or function. A function is written from the protein sequence and its fold, and these emergent function(s) are read by a network. Since the cell processes in the CDC are emergent properties of the complexes and networks, they directly control progression through the CDC, including the recursive process of DNA replication.

The functions, pathways, and cycles are also regulated by many different types of post-translational modifications, which are often added and removed during the life of a proteins, thus are short-term temporary memory and equivalent to Cache in a computer. Each of these cell processes that emerge from the Network are related to the execute function, thus are analogous to processor in computers.

**Do some steps in cycles mimic latches?** Memory in biological systems can be modeled using fundamental digital logic constructs. A simple memory circuit can be formed with two NOR gates or two NAND gates, each with two inputs and one output (**Fig. 3A**). In a NOR gate, when both inputs are off "0", the output is on "1"; in all other cases where one or both inputs are on, the output remains off. Most circuits constructed from such gates are combinational circuits, meaning their output depends solely on their current inputs.

In contrast, the SR latch represents a fundamentally different type of circuit—a sequential circuit—in which the output depends not only on the present inputs but also on the history of prior inputs, thereby creating a simple form of memory. An SR latch is constructed from two cross-coupled NOR gates, where the output of each gate becomes an input to the other (**Fig. 3B**). Inputs A and B act as external inputs, while outputs Q and Q̄ (the inverse of Q) encode the memory state.

There are three principal states in the SR latch:

> Set: When Input A is activated, Q is set to 1 and remains so until reset.
>
> Reset: When Input B is activated, Q is reset to 0 and maintained.
>
> Metastable (Forbidden) State: When both A and B are simultaneously active, the circuit enters a forbidden state where Q and Q̄ cycle continuously between 0 and 1. This unstable behavior results from continuous feedback between the gates, a dynamic unfamiliar to many biomedical scientists. A referenced video tutorial clarifies this unique behavior.[46]

Applying these digital logic principles to biological models, I examined one critical cycle within the Cellular Dogma Cycle (CDC): The Transcription Cycle, modeling its steps using logic gates, including memory elements (**Fig. 3C**).

**Initiation, Elongation, and Memory**. A transcription factor complex input from the Epigenetics Cycle and available ribonucleotide triphosphates (rNTPs) serve as dual inputs to an AND gate, producing pre-mRNA as the output. This is a progressive reaction, in which bases are sequentially added until the message is completed. This is an elongation and requires memory to continue adding bases is modeled with the output of the AND gate feeding back to both inputs (only feedback to one input is shown for simplicity). When one base is added, the output of 1 (active) is passed to the inputs and the transcription cycle is repeated again and again until a hairpin loop is encountered that causes disassociation of RNA polymerase. This maintains memory of the input as 1 until the termination signal is reset back to 0 upon encountering the hairpin. Since both inputs must be present, the AND gate logic accurately reflects the requirement.

**Splicing and Decision Making**. Once pre-mRNA is synthesized, it undergoes splicing to produce mature mRNA, a process requiring memory to persist until completion. This step is modeled by an SR latch, capturing the two mutually exclusive outcomes: 1. Translation (combination with ribosomes via an AND gate, leading to protein synthesis in the Protein Cycle), or 2. Degradation (combination with RNases via an AND gate, leading to Nonsense-mediated mRNA decay into rNMPs in the NMD Cycle). This mutual exclusivity can be modeled by paired inverse outputs in a SR latch in which the mRNA is either degraded or translated, never both—mirroring the forbidden state concept of the latch where simultaneous actions are not allowed.

**Recycling and Completion**. If the mRNA is to be translated, it enters the Protein Cycle. If the mRNA molecule is translated, it exits the Transcription Cycle and enters the Translation Cycle by combining with the ribosome. Both inputs are required with an AND gate to produce a protein chain output. Alternatively, if the mRNA molecule is degraded, in combines with an additional input of the RNase genes in the NMD cycle through an AND gate releases ribonucleotide monophosphates (rNMPs). Because both inputs are required for degradation, and complete degradation of an mRNA requires many cycles to remove bases one by one, this was modeled with an AND gate with output feedback to the input, similar to the earlier transcription step in the cycle.

The rNMPs substrates combine with polyphosphate kinase enzymes and ATP as inputs to regenerate rNTPs through an AND gate. As NTPs are regenerated, this completes the Transcription Cycle, as rNTPs are reused to make new RNAs.

Throughout these steps, memory is maintained by feedback mechanisms and logical constructs ensuring continuity until an explicit reset condition is met. Other cycles interact and are required to control the flow through the cycle.

Biological cycles may fundamentally operate using logic-based architectures, where feedback, progression, and decision-making are embedded forms of cellular memory distinct from, but complementary to information encoded in macromolecular sequences. The logic gate cycle construct of the Transcription Cycle may be similarly applicable to the previously listed cycles and may be a general way by which cell process and stores memory outside that of coding by macromolecules. For example, the Kreb's Cycle is also composed mostly of AND gates connected in a 10-step cycle. It also has several key decision points to continue anabolic use or redirect its metabolites to synthesize other amino acids and molecules through connected cycles; this could be modeled by a latch. These decision points allow the cell to either continue energy production or divert intermediates for biosynthetic purposes.

**The Central Dogma Cyclic Network.** The CDC is the main information flow cycle within the cell, but it is just one of many interlinked cycles that arise from functional protein networks. Each of these cycles connects with others—including the CDC—forming a dynamic, interconnected network. The output of one cycle often serves as the input for another, creating a web of cycle-to-cycle communication. This network encodes a distributed form of biological memory, collectively producing the protein complexes responsible for catalysis, regulation, and essential cellular processes. Each cellular behavior emerges from the activity of one or more cycles, which is a mechanism for memory of all cell process, alongside the inheritance of the gene sequences. As illustrated in **Figs. 2** and **3C**, these cycles are interconnected and likely represent evolutionary solutions to the need for cellular systems to remember and reproduce key processes reliably over generations.

A clear example of the breadth and importance of such cycles can be seen in signal transduction pathways. Many of these involve kinases, which act as "writers" by phosphorylating target proteins. These modifications are "read" by adaptor proteins like 14-3-3, or those containing SH2 or PTB domains, only when they are phosphorylated at a specific site(s). Subsequently, phosphatases function as "erasers" removing phosphate groups and restoring the protein to its original state.[15,48] During this phosphorylation cycle, the functional output of the modified protein often changes, influencing downstream processes. Given the diversity of kinases and phosphatases, as well as the numerous potential phosphorylation sites, thousands of such signaling-regulatory cycles likely operate simultaneously within a cell, each interacting with and modulating other cycles.

A parallel form of cyclical regulation governs the histone modification system and epigenetic memory. This system consists of enzymes that covalently modify histone tails (writers), proteins that recognize and bind these modifications (readers), and enzymes that remove them (erasers).[49–54] Increasing evidence suggests that cell signaling acts as a key input into this epigenetic cycle, influencing histone marks and triggering changes such as nucleosome disassembly and the initiation of the Transcription Cycle. Signals that alter these marks contribute to short-term epigenetic memory, shaping gene expression patterns over time.

The pathways described in this paper are presented in a simplified manner to demonstrate concepts. Each cycle presented is substantially more complex, with some additional steps, additional interactions with other cycles, and likely involve more sophisticated logical constructs. However, while subject matter experts may refine the logic and steps, the cyclic architecture, and therefore memory are the important concepts hypothesized herein.

**Discussion**

I present a model that extends the Central Dogma to the CDC and CDCN, emphasizing the role of interconnected molecular cycles in creating cellular memory. The DNA sequence alone is insufficient to establish the complex memories required for a cell to inherit function from its parent. Instead, emergent properties from proteins, functional pathways, molecular complexes, connections to other cycles, and the integration of small molecules are essential for maintaining cellular memory.

There are four distinct advantages of modeling the CDC and CDCN with cycle-driven memory when compared to the Central Dogma: 1. The CDC and CDCN models raise important avenues of inquiry; 2. They disambiguate and model cell reactions and processes with greater accuracy. 3. The framework is extensible, all possible 16 possible types of digital logic, better representing regulation and homeostasis; and 4. The CDC and CDCN frameworks can accommodate and describe the full complexity of biological information processing.

This model also reveals previously unaddressed questions. For example, how does a cell determine whether to degrade or translate an mRNA molecule? If degradation and translation occurred simultaneously, partially translated mRNAs would be destroyed, compromising protein synthesis. I hypothesize that this decision-making process is governed by a molecular regulatory circuit akin to an SR latch (**Fig. 3C**), a model that warrants experimental investigation. Notably, despite over 60 years of research, this critical regulatory decision has not been explicitly explored in the literature.

Current models and diagrams of signaling pathways and cellular processes are often overly simplistic, failing to capture the complexity of regulatory steps and decision points, as exemplified in the Transcription Cycle (**Fig. 3C**). Most biological pathway diagrams only depict basic logic gates such as OR, AND, and NOT. However, for systems with two inputs and one output, there are 16 possible digital logic operations. A more complete toolbox that embraces the full range of digital logic allows for the modeling of progressive reactions, timing mechanisms, and memory formation—all critical aspects of biological behavior. Recent work from my laboratory has shown that combinations of transcription factor mutations can collectively generate all possible logical outputs for transcription regulation supporting the use of logical encoding in nature.[17]

**Cyclic mechanisms of memory may be extensible to other systems.** This cycle model of memory is likely not limited to cells but may extend to other biological systems. It could serve as a broader framework for understanding how memory is created and managed across living systems.

Cycles are already well-recognized in physiological processes and may also underpin memory in cell-to-cell communication. A typical cell expresses approximately 300 different receptor types, forming the basis for extensive cyclic communication networks.[55] For example, in response to stress, the hypothalamic-pituitary-adrenal (HPA) axis operates as a feedback cycle: the hypothalamus secretes corticotropin-releasing hormone (CRH), stimulating the pituitary gland to release adrenocorticotropic hormone (ACTH), which in turn stimulates the adrenal glands to secrete cortisol. Cortisol then provides negative feedback to the hypothalamus, completing the cycle. This process establishes a form of memory—when the brain perceives stress, it initiates cortisol production to mobilize energy reserves and regulate blood pressure. Similarly, the formation and reactivation of neuronal ensembles may involve cyclic neuronal circuits that store and retrieve memories in animal brains.

Beyond individual organisms, ecosystems depend on the carbon cycle to maintain homeostasis between plants and animals. Plants capture sunlight energy to convert absorbed carbon dioxide into carbohydrates, releasing oxygen in the process. Animals consume these carbohydrates and oxygen to extract energy, returning carbon dioxide to the atmosphere through respiration. This cycle encodes a biological memory of how solar energy sustains ecosystems, supporting stability and balance.

From a medical perspective, disease pathology and treatment interventions can be understood as disruptions in cellular or physiological memory cycles. When these information-processing systems become dysregulated, cellular communication and homeostasis deteriorate, ultimately leading to disease. Viewing disease through the lens of biological memory cycles offers new insights into therapeutic strategies focused on restoring normal information flow and cellular function.

**Figures**

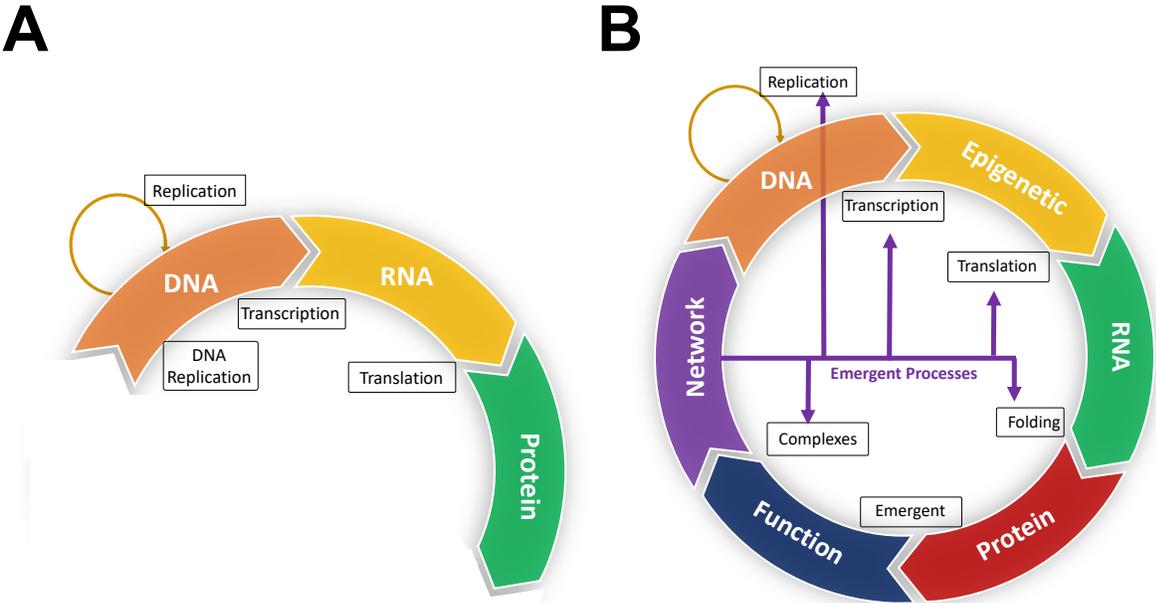

**Figure 1.** *The Central Dogma Cycle*. **A.** *Central Dogma* pathway for genetic information flow in cells. DNA is replicated or transcribed into RNA, which is next translated into proteins. **B.** The proposed *CDC* for information flow in cells; 3 new steps are added to the Central Dogma completing the cycle. An epigenetic step is included between DNA and RNA transcription. After translation, protein fold producing molecular function(s). Proteins form complexes with other proteins, other macromolecules, and small molecules to create cell processes. Each emergent cell processes (purple arrows) forms a separate cycle that produce emergent cell processes (purple arrows). The figure was created with PowerPoint.

# Central Dogma Cyclic Network

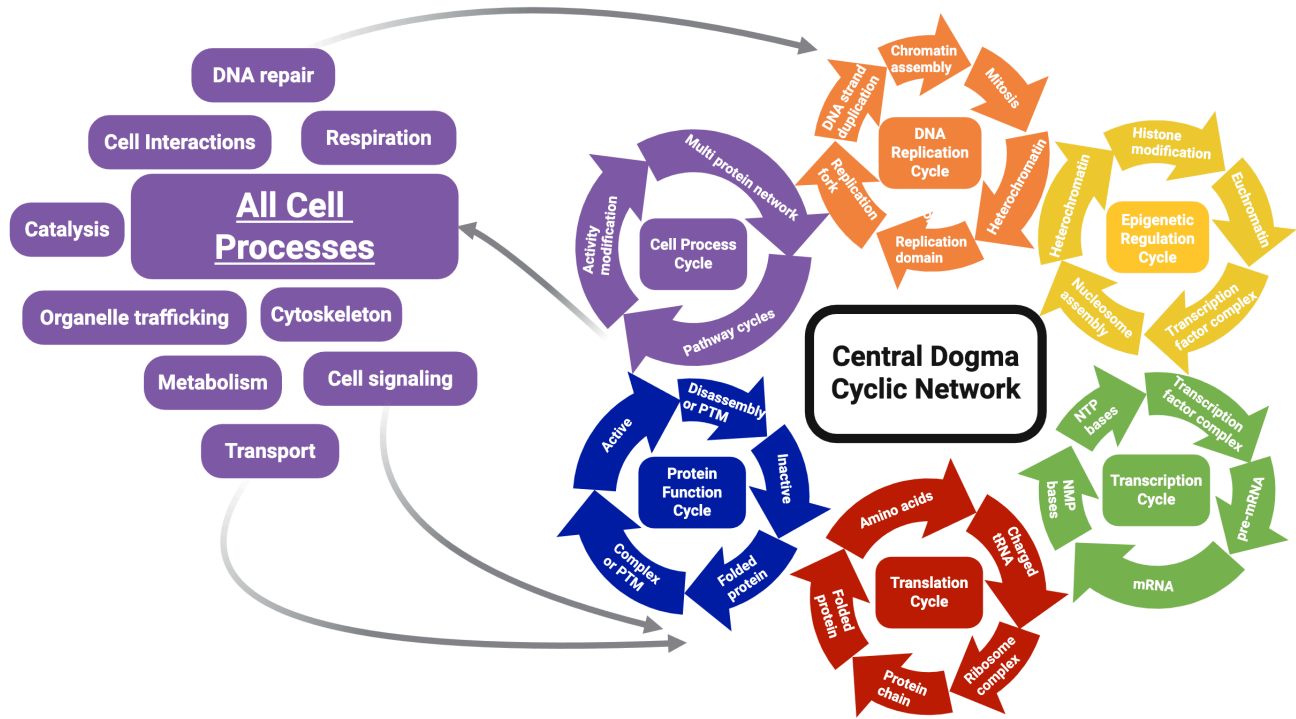

**Figure 2. Central Dogma Cycle and Central Dogma Cyclic Network model. A.** CDC from **Figure 1** expanded to show steps for an interconnected series of cycles. The colors of the cycles match the steps in **Figure 1B.** The output of the CDC produces many other cycles for cell for numerous cell processes (some are shown in purple boxes). These interconnected cycles are called the Central Dogma Cyclic Network (CDCN), which also includes the CDC as a series of core cycles in the CDCN. The figure were generated with BioRender.

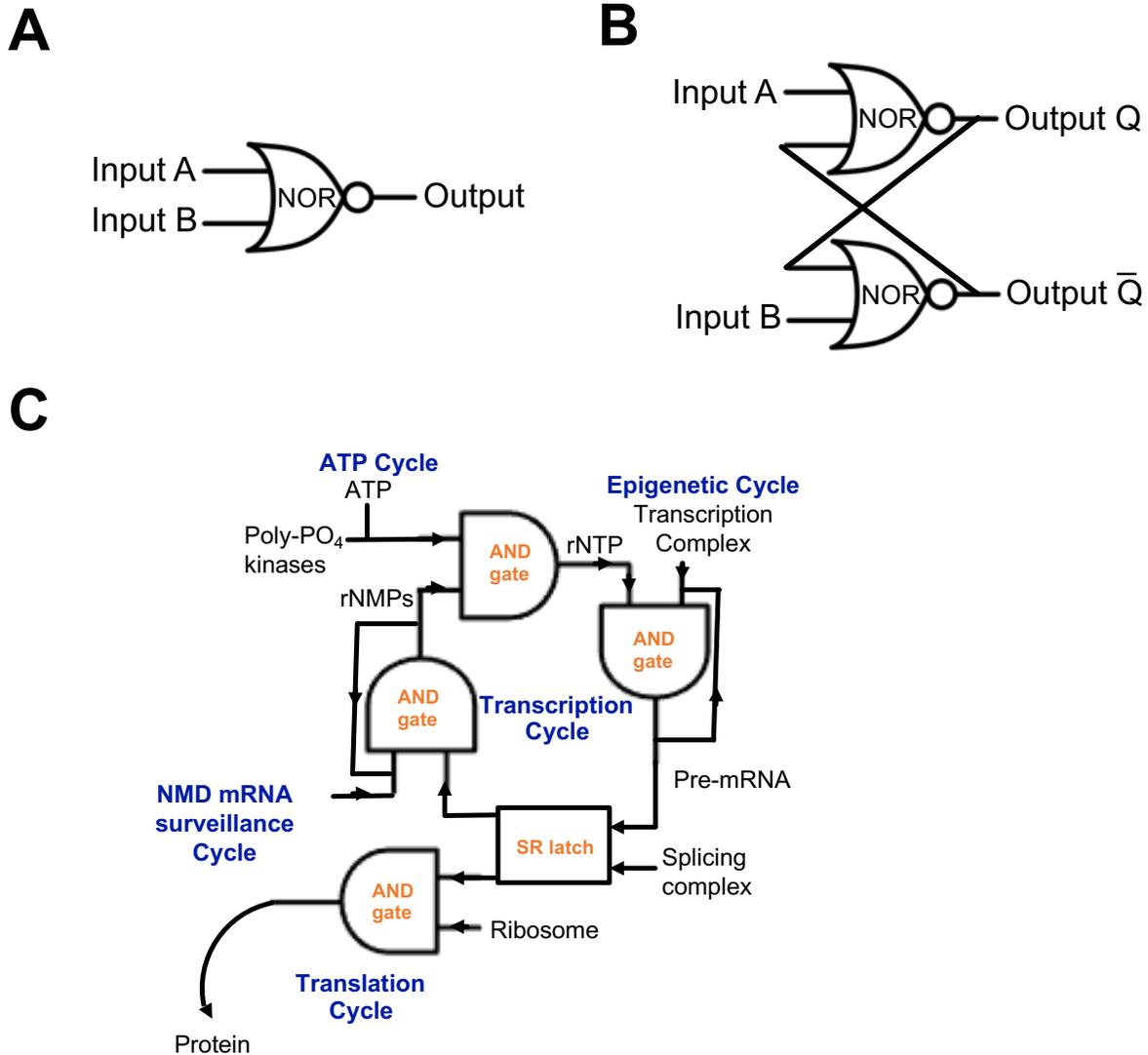

**Figure 3. Memory encoding in computers and applied to the Central Dogma Cycle. A.** A NOR gate only produces an output when both inputs are off. **B.** One version of a *bit* (SR latch) is built from a circuit combining two NOR gates, each with two inputs and two outputs (A and B). One output of each NOR gate is one of the two inputs of the other NOR gate with Inputs A and B being the additional inputs for each gate. Output Q and its inverse, Q̄ are the outputs of each gate. 0 is considered no input and 1 is considered and input. The output behaviors of each input are dependent on the history of the other input. **C.** An example expressing how the Transcription Cycle can be encoded by digital logic gates to create memory and decision making. Cycles are labeled with blue fonts and electrical symbols with orange fonts. Four steps of the Transcription cycle (transcription, pre-mRNA splicing, RNA degradation, and RNA base regeneration), one step inputting to the Translation cycle in the CDC, and the Nonsense Mediated Decay (NMD) and ATP cycles from the broader CDCN are shown. The figure were generated with PowerPoint.

**Potential Conflict of Interest**

I disclose a potential conflict of interest as Martin R. Schiller is both a Professor at the University of Nevada Las Vegas and an employee of Heligenics Inc., which produces and sells Mutation Effect on Gene Activity (MEGA)-Maps produced with the GigaAssay.

**Acknowledgements**

I wish to thank Lancer Brown for discussions related to the ideas in the paper and for his help in creating and modifying figures. The development of the concepts for this work was supported by grants from the National Institutes of Health (NIH): R21AI116411, P20GM121325, and R44AI188872.